\documentclass[twocolumn,showpacs,prl,superscriptaddress]{revtex4}

\usepackage{amsmath}
\usepackage{bm}
\usepackage{array}
\usepackage{graphicx}
\usepackage{dcolumn}   
\usepackage{amssymb}
\usepackage[english]{babel}

\renewcommand{\matrix}[1]{\begin{pmatrix} #1 \end{pmatrix}}
\newcommand{\onrt}{\frac{1}{\sqrt{2}}} 

\newcommand{\halfof}[1]{\frac{#1}{2}}

\newcommand{\ket}[1]{| #1\rangle}

\newcommand{\p}{\phi}
\renewcommand{\t}{\theta}

\begin{document}


\title{Nanoscale magnetometry through quantum control of nitrogen-vacancy centres in rotationally diffusing nanodiamonds}
\date{\today}

\author{D. Maclaurin}
\affiliation{School of Physics, The University of Melbourne, Parkville,
3010, Australia}
\affiliation{Department of Physics, Harvard University, Cambridge, MA 02138, USA}
\author{L.T. Hall}
\affiliation{School of Physics, The University of Melbourne, Parkville,
3010, Australia}
\affiliation{Centre for Quantum Computation and Communication Technology, School of Physics, The University of Melbourne, Parkville 3010, Australia}
\author{A.M. Martin}
\affiliation{School of Physics, The University of Melbourne, Parkville,
3010, Australia}
\author{L.C.L. Hollenberg}
\affiliation{School of Physics, The University of Melbourne, Parkville,
3010, Australia}
\affiliation{Centre for Quantum Computation and Communication Technology, School of Physics, The University of Melbourne, Parkville 3010, Australia}

\begin{abstract}
The confluence of quantum physics and biology is driving a new generation of quantum-based sensing and imaging technology capable of harnessing the power of quantum effects to provide tools to understand the fundamental processes of life. One of the most promising systems in this area is the nitrogen-vacancy centre in diamond - a natural spin qubit which remarkably has all the right attributes for nanoscale sensing in ambient biological conditions. Typically the nitrogen-vacancy qubits are fixed in tightly controlled/isolated experimental conditions. In this work quantum control principles of nitrogen-vacancy magnetometry are developed for a randomly diffusing diamond nanocrystal. We find that the accumulation of geometric phases, due to the rotation of the nanodiamond plays a crucial role in the application of a diffusing nanodiamond as a bio-label and magnetometer. Specifically, we show that a freely diffusing nanodiamond can offer real-time information about local magnetic fields and its own rotational behaviour, beyond continuous optically detected magnetic resonance monitoring, in parallel with operation as a fluorescent biomarker.

\end{abstract}
\maketitle

\section{Introduction}

The diamond nitrogen-vacancy (NV) defect shows great promise as a biological imaging tool in two distinct capacities. Firstly, its optical properties and biocompatibility make it an excellent fluorescent biolabel \cite{functionalised_surface,low_toxicity,Cheng,PNAS_biolabels,cancer_drugs,neugart,Fann,Treussart}. It is extremely bright, highly photostable \cite{photostable_bleach}, and it can be imaged with spectacular spatial resolution \cite{Bradac10} through sub-diffraction localization techniques like stimulated emission depletion  \cite{STED}. 
Secondly, optical spin polarisation and measurement and microwave control, the NV defect's electron spin can be used to detect magnetic fields with nT Hz$^{-1/2}$ \cite{Harvard_exp,Nature_mat} sensitivity, with potentially nano-metre spatial resolution \cite{Stuttgart_exp,Maletinsky10}. The prospect of using the NV centre for nanoscopic MRI has significant promise for a range of problems in biology, including: detecting operation of individual neuronal ion channels \cite{ion_channels} and atom-scale magnetic resonance imaging \cite{Taylor,Liam2}.

However, the more revolutionary application is the combination of the NV as a nano-scale biomarker and magnetometer. This would be a super-optical probe capable of reporting motional information simultaneously with atomic level magnetic fields and their fluctuations over arbitrary long times.  Such a  system would be able to give the precise location of some attached ligand as it moves through a cell, as well as information about its molecular environment by detecting nanoscale magnetic fields. 
However, nearly all demonstrations of NV magnetometry have assumed the NV spin quantization axis is effectively stationary with respect to the lab-frame control fields. In order to employ the NV centre as a combined nanoscale bio-marker/magnetometer one must fully understand the effect of random rotational motion on the nanomagnetometry and control protocols. In  Ref. \cite{Scholten} NV nanodiamond nanomagnetometry in living cells was demonstrated. In these experiments, the measurements were conducted on a endocytosed nanodiamond which moderated the motional timescales, facilitating quantum measurement and allowing information about the orientation to be determined over long periods.   In this paper we explore the quantum control of a rotationally diffusing NV system and find that the accumulation of geometric phases, due to rotation 
\cite{Dima_Book,Dougal_Thesis,AC_paper,Berry_NV,Dima,Paola}, is the critical phenomenon in the application of the NV centre as a bio-marker/magnetometer. 


\begin{figure}
\centering
\includegraphics[width=8.5cm]{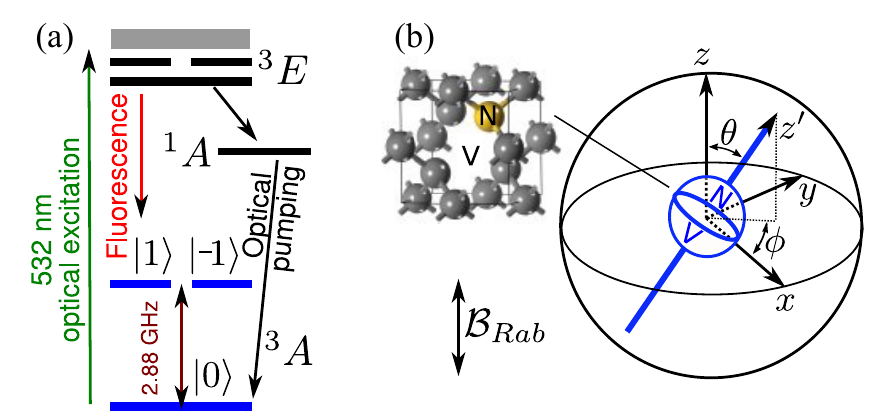}
\caption{(a) Energy level diagram of the NV center. (b) Geometry. The magnetic field direction of the Rabi pulse defines the $z$-axis. $z'$ is the instantaneous direction of the N-V axis, defined with respect to the lab-frame unprimed coordinate system by $\theta$, $\phi$.  \label{one}}
\vspace{-0.5cm}
\end{figure}

\section{The diamond NV system}

The NV defect has a spin triplet ground state with a 2.88 GHz zero-field splitting between the $m=0$ state and the degenerate $m=\pm 1$ states as shown in Fig. 1(a). Optical excitation at 532 nm  can pump the defect into the $m=0$ state, and allows the population of the ground state to be read, since the $m=0$ state has a higher fluorescence than the $m=\pm 1$ states. The effective Hamiltonian of the NV ground state triplet in the presence of a magnetic field $\bm B$ [lab frame coordinate system defined by $z$, see Fig 1(b)], ignoring crystal asymmetries and hyperfine effects, is
\begin{equation} H = \frac{1}{\hbar}D S_{z'}^2 +\gamma \bm B \cdot \bm S, \label{Hamiltonian} \end{equation}
where $\gamma$ is the hyro-magnetic ratio of the NV ($\gamma=g\mu_{\rm B}/\hbar$, $g\approx 2$). The first term is the zero-field splitting of the NV system itself, where $D$ is the zero-field splitting strength. It is this term which makes the crystal's orientation crucial. It defines a quantisation direction $z'$  [see Fig. 1(b)] which lies along the axis connecting the nitrogen atom to its adjacent vacancy. The second term is the usual Zeeman splitting interaction with a magnetic field. 

Previously it has been shown that the NV centre can be used to measure DC and AC magnetic fields \cite{Harvard_exp, Stuttgart_exp} and fluctuating (FC) fields \cite{Liam,Decoherence} and rotation \cite{Scholten,Dima_Book,Dougal_Thesis,Berry_NV,Dima,Paola}. These protocols work by optically polarizing the NV $m=0$ spin state. A $\pi/2$ microwave pulse then establishes a coherent superposition of the $m=0$ and $m=1$ states, and the system evolves under the influence of any magnetic fields by the Zeeman effect. After an evolution time, $\tau$, the rotation and strength of the magnetic field manifests itself as a phase, relative to $m=0$, of the $m=1$ state. In the adiabatic limit this phase is 
\begin{eqnarray}
\Phi=\Phi^{\rm G}+\Phi^{\rm D}=\int_P \cos \theta d\phi+\gamma\int_0^{\tau} {\hat z}^{\prime}(t)\cdot {\bf B}(t) dt, \nonumber \\ 
\end{eqnarray}
where $\Phi^{\rm G}$ is the accumulated geometric phase, due to rotation along some path, $P$, defined by $\theta$ and $\phi$ [see Fig. 1(b)] \cite{Dougal_Thesis,Berry_NV} and $ \Phi^{\rm D}$ is the accumulated dynamical phase due to the change in the projection of the magnetic field, $\bm B$ onto the NV axis, ${\hat z}^{\prime}$. A second $\pi/2$ pulse converts this phase into a population difference, which is read out directly as a change in fluorescence. 

It is common to apply a static magnetic field to split the $m = \pm 1$ states, selecting only one of the $m = \pm 1$ states by using a sufficiently weak microwave field resonant with one of the transitions (Rabi field). Since a static field would cause different splitting depending on the crystal's orientation \cite{Scholten}, we do not take this approach. Instead, we assume that the degeneracy of the $m = \pm 1$ states is lifted by strain, which is usually the case for nanodiamonds \cite{Scholten}. 

\section{Model of Diffusion}
\subsection{Motion of the crystal}

\begin{figure}
\centering
\includegraphics[width=8.5cm]{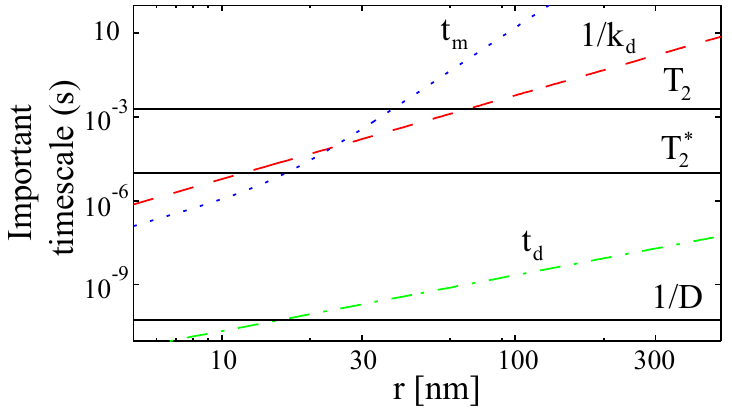}
\caption{Important timescales as a function of crystal radius, assuming the crystal is immersed in room  temperature water ($T=300$ K) with a viscosity of $\eta=10^{-3}$ Pa s. Red dashed curve: characteristic rotational diffusion timescale, $1/k_d$. Blue dotted curve: population mixing timescale, $t_m$, for non adiabatic motion. Green dashed-dotted curve: angular velocity damping timescale, $t_d$. Also shown are the typical timescales associated with the NV spin state: homogeneous broadening ($T_2$), inhomogeneous broadening ($T_2^*$) and zero field splitting ($1/D$). \label{time_scales}}
\vspace{-0.5cm}
\end{figure}

A spherical crystal is considered, whose rotation follows Brownian motion, according to the Langevin equation
\begin{equation}
I\dot{\bm \omega} = -\gamma_d \bm \omega + \bm \tau(t),\label{langevin}
\end{equation}
where $I$ is the crystal's moment of intertia, $\bm \omega$ is the crystal's angular velocity, $\bm \tau$ is a stochastic delta function correlated torque, and $\gamma_d$ is a drag coefficient given by Stokes' law, $\gamma_d = 8\pi\eta r^3$, where $r$ is the crystal's radius and $\eta$ is the fluid's viscosity. It can be imagined that the crystal rotates in a `random walk' of discrete small angle steps, whose directions are uncorrelated with one another. This rotational motion is characterised by a single diffusion coefficient $k_d$ which, from Eq. (\ref{langevin}) and the equipartition theorem, is given by
\begin{equation}
k_d = \frac{k_B T}{8 \pi r^3 \eta},
\end{equation}
where $k_BT$ is the thermal energy. Thus, in room-temperature water, $r = 5$ nm [$r = 50$ nm] gives $k_d \approx 1$ rad$^2$/$ \mu$s [$k_d\approx 1$ rad$^2$/ms], see red dashed curve in  Fig.~\ref{time_scales}. 

\subsection{Ensemble averaged measurements}
In practice, and even in theory, a single measurement, of a quantum system, contains very little information. The quantity which is measured  is a time averaged signal from a large ensemble of individual measurement runs. Environmental  perturbations cause subtly different quantum evolution in each member of the ensemble, leading to a loss of information in an ensemble-averaged signal after a certain evolution time. In the case of the rotationally diffusing nanodiamond, the ensemble's members differ more dramatically. In addition to its own microscopic environment each member of the ensemble has its own starting orientation and rotational trajectory.

A series of sequential measurements is made which may involve several NV centres in a single crystal. If the time over which measurements are made is sufficiently small compared with the rotational diffusion time, $1/k_d$, then the situation would be comparable to the case of a static crystal. In this work the opposite limit is considered and it is assumed that sufficiently many measurements are made that the final signal is an average over the full theoretical ensemble: all initial orientations and rotational trajectories are accounted for. Experimentally this average can be constructed via the repeated measurement of the same NV centre or over a collection NV centres \cite{Scholten}.

For such a scenario consider a continuously applied microwave field, of amplitude $B_R$, tuned to the 
$\psi_{z^{\prime}}^{(m=0)} \rightarrow \psi_{z^{\prime}}^{(m=1)}$ transition. This will cause Rabi oscillations with a frequency which depends on the 
angle, $\theta$, between the NV axis $z^{\prime}$, and $z$, the oscillation direction of the microwave magnetic 
field. Initialised into 
$\psi_{z^{\prime}}^{(0)} $, the population of 
$\psi_{z^{\prime}}^{(0)} $ for a single NV as a function of time is 
then 
\begin{eqnarray}
P_0 = [1+\cos(\Omega_Rt \sin \theta)]/2,
\end{eqnarray}
where $\Omega_R = g\mu_B B_R/\hbar$ is the Rabi frequency when $\theta = \pi/2$. Assuming $\Omega_R \gg 1/k_d$ the ensemble-averaged signal is an average 
over all the possible different Rabi frequencies:
\begin{eqnarray}
S_0(t)&=&\frac{1}{4}\int_0^{\pi} d\theta \sin \theta \left[1+\cos \left(\Omega_R t \sin \theta\right)\right] \nonumber \\
&=& \frac{1}{4}\left[2+\pi H_{-1} \left(\Omega_Rt\right)\right],
\end{eqnarray}
where $H_{\alpha}$ is the Struve function. $S_0(t)$ (solid curve) is shown in Fig.~3(a), along with the equivalent curve for a static crystal (dashed curve), with $\theta=\pi/2$. For an ensemble averaged measurement  the signal reaches its first minimum at $t \Omega_R \approx 1.16 \pi$  at which point $S_0 \approx 0.14$.

This result enables the determination of the optimal microwave pulse time to achieve a $\pi/2$ rotation. The required time is simply read off from the curve. In  our system, there is a  certain amount of freedom in deciding how long a $\pi/2$ pulse should be. We could take it to be $t_{\pi/2} = \pi/(2\Omega_R)$, which would produce a $\pi/2$ rotation for an optimally aligned ($\theta = \pi/2$) NV centre, and a smaller rotation otherwise. Alternatively, we could use $t_{\pi/2} = 1.16\pi/(2\Omega_R)$ which, based on Fig. 3(a), would optimise the initial amplitude of a Ramsey-type experiment. One of the strengths of Ramsey-type experiments is their robustness to the $\pi/2$ pulse time chosen. We do not, therefore, expect this decision to be an important one (as simulations will confirm). For now we keep our discussion general, introducing a parameter, $a$, to describe the pulse time. As such when we refer to a `$\pi/2$' pulse we mean a pulse of length $t_{\pi/2} = a\pi/(2\Omega_R)$, where $a \approx 1$, which produces a spin rotation of  $\Theta = a(\pi/2) \sin \theta$.

Now consider a Ramsey-type pulse sequence. The NV centre, starting with $z^{\prime}$ at some polar angle $\theta_1$ to the $z$-axis, is first optically pumped into 
$\psi_{z^{\prime}}^{(0)} $. A `$\pi/2$' pulse is then 
applied, tuned to the 
$\psi_{z^{\prime}}^{(0)}  \rightarrow \psi_{z^{\prime}}^{(1)} $ transition, producing a spin rotation of $\Theta_1= a(\pi/2) \sin \theta_1$. 
The system is allowed to evolve for time $\tau$, during which time the crystal rotates through 
some trajectory and the 
$\psi_{z^{\prime}}^{(1)} $ state develops a phase $\Phi$, due to rotation and the presence of magnetic fields. A final `$\pi/2$' 
pulse is applied. This final pulse may be in phase with the initial pulse, or out of phase 
by $\pi$. It produces a rotation of 
$\pm \Theta_2 \equiv \pm a(\pi/2) \sin \theta_2$ where $\theta_2$ is the final angle between $z^{\prime}$ and $z$ and the $+$ ($-$) refers to the pulse being in (out) of phase with the original 
pulse. The final population $P_0$ of the 
$\psi_{z^{\prime}}^{(0)} $ state can then be written down 
explicitly as
\begin{eqnarray}
P_0=\frac{1}{2}\left[1+\cos \Theta_1 \cos \Theta_2 \pm \sin\Theta_1 \sin \Theta_2 \cos \Phi\right].
\label{Probability}
\end{eqnarray}
The final, normalised, signal is the ensemble average of this value $S(t)=\langle P_0\rangle$, where the ensemble average is taken over all possible starting orientations and rotational trajectories.  The upper and lower 
bounds of the {\it signal envelope} are produced by the two curves implied by the $\pm$ term 
of Eq.~(\ref{Probability}), provided the mean phase vanishes, 
$\langle \Phi\rangle=0$ (true for the cases considered in this work). This work primarily focuses on the behavior (typically decay) of the signal envelope, due to rotation and the presence of magnetic fields, rather than the oscillations themselves. Physically, the two curves correspond to the 
phase ($0$ or $\pi$ respectively) of the final `$\pi/2$'  pulse.

\begin{figure}
\centering
\includegraphics[width=8.5cm]{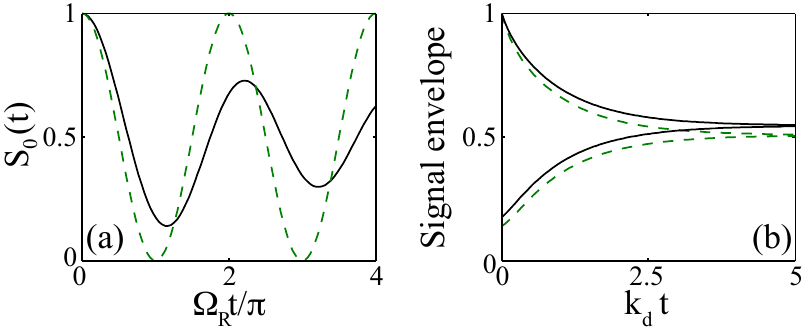}
\caption{(a) Rabi  signal  averaged over the ensemble of possible crystal orientations (solid black curve) and for optimal crystal orientation ($\theta = \pi/2$) (dashed green curve). (b) Signal envelope showing exponential decoherence due to geometric phase accumulation, for  $a = 1$ (solid curve) and $a = 1.16$ (dashed green curve). \label{one}}
\vspace{-0.5cm}
\end{figure}



\section{Dephasing due to geometric phase accumulation}

To see the influence of the geometric phase, we consider how the spin state of the rotating crystal evolves in the absence of external fields. As the crystal rotates, the direction of $z'$ changes, so that Eq.~(\ref{Hamiltonian}) becomes a time-varying Hamiltonian. With respect to the eigensates of spin projection along the fixed $z$ axis, the zero-field Hamiltonian is:
\begin{eqnarray}
& &H_0=\frac{1}{\hbar}D S_{z'}^2 = \nonumber \\ & &D\hbar\left(
\begin{array}{ccc}
  \cos^2\theta +\frac{\sin^2\theta}{2} & \frac{e^{-i\phi}\cos\theta\sin\theta}{\sqrt{2}}& \frac{e^{-2i\phi}\sin^2 \theta}{2} \\
\frac{e^{i\phi}\cos\theta\sin\theta}{\sqrt{2}} & \sin^2\theta & -\frac{e^{-i\phi}\cos\theta\sin\theta}{\sqrt{2}} \\
 \frac{e^{2i\phi}\sin^2 \theta}{2}   & - \frac{e^{i\phi}\cos\theta\sin\theta}{\sqrt{2}} &  \cos^2\theta +\frac{\sin^2\theta}{2}
\end{array}
\right)_z,
\nonumber \\
\end{eqnarray}
where $\theta$ and $\phi$ are time dependent. If the rate of rotation is small compared with $2.88$ GHz the evolution will be adiabatic, producing no changes in populations of the spin sublevels (with respect to the $z'$ quantisation axis). Each sublevel, however, develops a geometric phase
\begin{equation}
\Phi_m=\int_0^t \left[\frac{d \psi_{z^{\prime}}^{(m)}(t)}{dt} \right]^{*} \psi_{z^{\prime}}^{(m)}(t) dt.
\end{equation}
There is a gauge degree of freedom in defining the phase of the eigenstates \cite{Dougal_Thesis,Berry_NV}. We choose to define the eigenstates with respect to the fixed $z$ basis as
\begin{eqnarray} 
\psi_{z^{\prime}}^{(1)}&=&\matrix{e^{-i\p}\cos^2\halfof{\t}\\ \onrt \sin\t \\ e^{i\p}\sin^2\halfof{\t}}_z , \,\,\,\,\, 
\psi_{z^{\prime}}^{(0)}=\matrix{-\onrt e^{-i\p}\sin\t \\ \cos\t   \\ \onrt e^{i\p}\sin\t}_z, \nonumber \\ 
\psi_{z^{\prime}}^{(-1)}&=&\matrix{e^{-i\p}\sin^2\halfof{\t}\\ -\onrt \sin\t \\ e^{i\p}\cos^2\halfof{\t}}_z.\nonumber \end{eqnarray}
This convenient choice of gauge means that an explicit phase factor between sublevels corresponds to the absolute phase of a microwave field linearly polarized along the $z$ axis. The relative geometric phase accumulation, between $\psi_{z^{\prime}}^{(0)}$ and  $\psi_{z^{\prime}}^{(1)}$, due to the crystal's rotation then evolves according to $\dot  \Phi = \dot \phi \cos \theta$,
where the phase evolution of each state is given by ${\dot \Phi}_m=m{\dot \Phi}$ \cite{Dima_Book,Dougal_Thesis,Berry_NV,Dima,Paola}.

Given a particular initial orientation 
$\theta_1$ the probability density of a particular final orientation $\theta_2$ and geometric
phase $\Phi$ after time $t$ is $p(\theta_2 , \Phi; t 
|\theta_1)$. The ensemble-averaged signal can then 
be expressed as the integral of the final ground state 
population over all possible $\theta_1$, 
$\theta_2$ 
and $\Phi$, weighted by the probability distribution function and the probability of the NV starting with orientation $\theta_1$:
\begin{eqnarray}
S(t)&=&\frac{1}{2}\int d \theta_1 d \theta_2 d \Phi \sin \theta_1 p\left(\theta_2 , \Phi; t 
|\theta_1\right) P_0\left(\theta_1,\theta_2,\Phi \right) \label{envelope}, \nonumber \\
&&
\end{eqnarray}
where $P_0(\theta_1,\theta_2,\Phi)$ is given by Eq.~(\ref{Probability}). The Fokker-Planck equation for the evolution of $p\left(\theta_2 , \Phi; t 
|\theta_1\right)$ is
\begin{eqnarray}\frac{1}{k_d} \frac{d}{dt} p(\t_2,\Phi,t|\t_1) =&-& \frac{\partial}{\partial \t_2}\left [ \sin \t_2 \frac{\partial}{\partial \t_2}\left ( \frac{p\left(\theta_2 , \Phi; t 
|\theta_1\right)}{\sin\t_2}\right )\right ] \nonumber \\
&+& \cot^2\t_2 \frac{\partial^2}{\partial \Phi^2}p\left(\theta_2 , \Phi; t 
|\theta_1\right). \label{FP}\end{eqnarray}
The first term describes the probabilistic rotation of the crystal, and the second term describes phase evolution, which deterministically depends on the probabilistic evolution of azimuthal angle $\p$. The initial condition is
\begin{equation}
p(\t_2,\Phi,t=0|\t_1) = \delta(\Phi)\delta(\t_2 - \t_1),\end{equation}
and the boundary conditions are periodic in $\theta$ and $\Phi$.


Numerically solving Eq. (\ref{FP}) gives the solution shown in Fig.~3(b), demonstrating that the decaying envelope is relatively insensitive to the `$\pi/2$' pulse duration, $a$. There is approximately exponential decay of the signal envelope over the timescale of the crystal's rotation, with decay time $\approx 0.90/k_d$. The diffusion rate, $k_d$, thus sets an upper bound on the coherence time for the crystal. In the absence of any external fields, measurement of this decoherence rate allows measurement of $k_d$ with shot noise-limited sensitivity $\Delta k_d \sqrt T \approx \frac{1}{\alpha \sqrt{N \tau_d}}$, where $\tau_d$ is the overall decoherence time ($1/\tau_d \approx 1/T_2^* + k_d$),  $\alpha\approx 0.01$ is a parameter which accounts for the non-zero fluorescence of the $m=\pm 1$ states, and finite collection efficiency \cite{Taylor}. $N$ is the number of NV centres in the crystal, along all four crystallographic axes, and $T$ is the total measurement time. Figure 4 (red dashed curve) shows the sensitivity as a function of crystal radius, with an assumed NV density of $10^{16}\text{cm}^{-3}$.

\begin{figure}
\centering
\includegraphics[width=8.5cm]{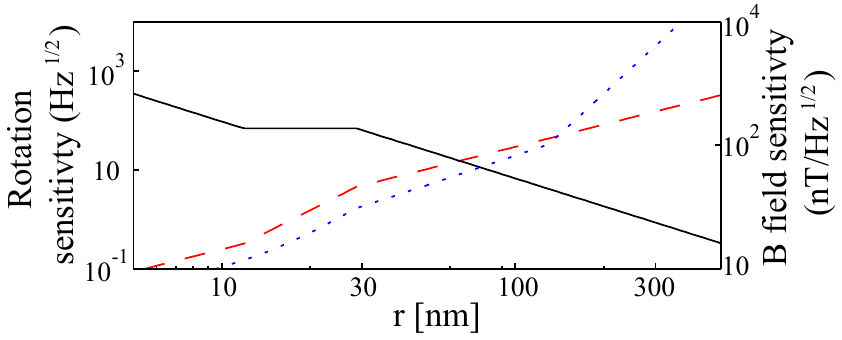}
\caption{Rotation rate and DC magnetic field sensitivities, for an assumed fluorescence collection efficiency of  $\alpha = 0.01$ and a density of NV centres in the crystal of $10^{16}$cm$^{-3}$. Solid black  curve, right axis: DC magnetic field sensitivity. Red dashed curve, left axis: rotation rate sensitivity from geometric phase decoherence. Blue dotted curve, left axis: rotation rate sensitivity from non-adiabatic population mixing. \label{sensitivity}}
\vspace{-0.5cm}
\end{figure}

\section{Dephasing due to population mixing with non-adiabatic evolution}
Up to this point we have been working within the adiabatic approximation. In the context of
the NV centre, the adiabatic criterion is
\begin{eqnarray}
\frac{1}{2}\left(\dot{\theta}^2+\dot{\phi}^2\sin^2\theta\right) \ll D^2.
\end{eqnarray}
The angular velocity of the nanodiamond is a consequence of thermal motion. Using the
equipartition theorem, the adiabatic criterion for the rotationally diffusing nanodiamond
is
\begin{eqnarray}
D\gg\frac{k_BT}{I}=\frac{15k_BT}{8\pi \rho r^3}=15 k_d,
\end{eqnarray}
where $\rho$ is the density of diamond. At room temperature this criterion will be satisfied if the crystal radius is much greater than $0.2$ nm [see Fig. 2, dashed red curve].

An additional adiabatic criterion, often overlooked, is that the state of the NV centre changes slowly compared to $1/D$. The timescale on which the NV state changes is the angular velocity damping time, $t_d=I/\gamma_d=\rho r^2/(15\eta)$. In Fig. 2(a) $t_d$ (green dashed-dotted curve) is plotted as a function of the radius of the crystal, for a crystal in room temperature water. From this it can be seen that   $t_d$ is smaller than $1/D$ when the crystal radius is smaller than $20$ nm.   

The geometric phase accumulation comes from a zeroth order perturbative expansion of the time evolution operator in the ratio of the rotation rate to the $2.88$ GHz zero field splitting. Taking the expansion to higher orders enables the population mixing that comes from nearly adiabatic evolution to be evaluated.
To evaluate the properties of an ensemble measurement consider an arbitrary initial state of the NV centre, where each state has a probability weighting of $P_{m}$ and phase $\Phi_m$. As shown in Appendix A the ensemble average population mixing rate is
\begin{eqnarray}
\langle k \rangle=\frac{2k_d}{1+t_d^2D^2}.
\end{eqnarray}
The above provides an expression for the population mixing rate in terms of physical parameters: the zero-field splitting frequency $D$, the rotational diffusion rate $k_d$ and the angular velocity damping time $t_d$.

To obtain the above results several assumptions have been made. Most significantly, the diffusive rotation of the nanodiamond  has been modeled as a series of discrete steps, using
physical arguments to infer the appropriate statistical behaviour of the steps' sizes and
durations. To check the validity of these approximations a numerical
Monte Carlo simulation has also been employed. The simulation models the Langevin equation (\ref{langevin}), using a torque which fluctuates on a finite but very small timescale, to produce a set of rotational trajectories. For each trajectory the Schr{\"o}dinger equation is solved numerically. The
results, for an NV centre which starts in $\psi_{z^{\prime}}^{(0)}$, are plotted in Fig. 5(a). They show excellent agreement with our analytical expression:
 \begin{equation}
S_m(t) = \frac{1}{3} + \frac{2}{3}\exp(-t/t_m). \label{non_adiabatic_decay}
\end{equation}
where $t_m=1/(3\langle k \rangle)$ is the population mixing time.

The implication of this non-adiabatic behaviour is more than merely a correction
to the behaviour expected from geometric phase accumulation. Measuring the population mixing rate
(effectively measuring $T_1$) provides another way to gauge the rotational diffusion rate. More over the measurement can be achieved without microwave fields. It is enough to simply initialise the NV into the $\psi_{z^{\prime}}^{(0)}$ state and observe the decay of the fluorescence rate as a function of free evolution time. The blue dotted curves in  Figs. 2 and 4  show the population mixing time, and associated sensitivity, respectively, as a function of the crystal radii.

\begin{figure}
\centering
\includegraphics[width=8.5cm]{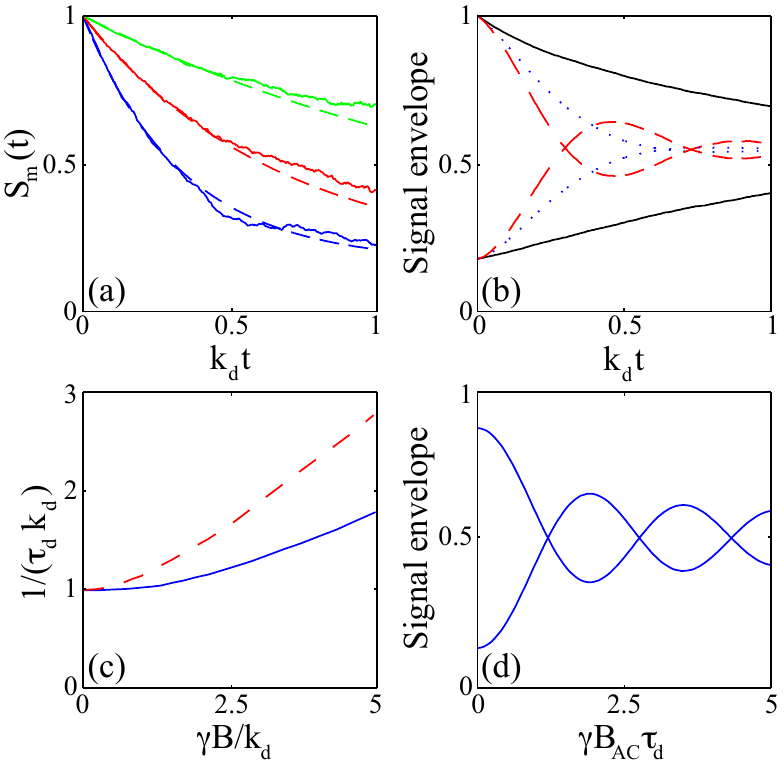}
\caption{(a) Population mixing due to non-adiabatic evolution: expected fluorescence signals as a function of free evolution time, relative to rotational difusion time, given initial occupation of $\ket 0$. The dashed curves follow the theoretical exponential decay of Eq. (\ref{non_adiabatic_decay}) while the solid curves are based on a Monte Carlo simulation of the Langevin equation followed by numerical solution of Schrodinger's equation. Blue $t_dD = 0.1$, red: $t_dD = 1$, green: $t_dD = 2$. (b)  Decoherence curves for several DC field strengths and orientations. Solid: no field. Dashed: $B = 10 k_d/\gamma \hat z$. Dotted:  $B = 10 k_d/\gamma \hat x$. (c) Decoherence rate relative to $k_d$ for DC magnetic fields in $z$ (solid) and $x$ (dashed) directions.  (d) Decoherence signal as a function of field strength for a spin-echo pulse sequence with an AC field in the limit of slow rotation compared to free evolution time. \label{one}}
\vspace{-0.5cm}
\end{figure}

\section{Static and fluctuating magnetic fields}
We now investigate the properties of rotationally diffusing NV centres in the presence of static (DC), fluctuating (FC) and oscillating (AC) magnetic fields. Magnetic fields enter our model by advancing the phase of the NV centre's spin sublevels. Under the secular approximation Eq. (\ref{Hamiltonian}) becomes
\begin{eqnarray}
H \approx \frac{1}{\hbar}D S_{z'}^2 + \gamma(B_z\cos\t + B_x\sin\t\cos\p), 
\end{eqnarray}
 it is only the field component parallel to $z^{\prime}$ (the NV axis) which affects the phase. 

\subsection{Static magnetic fields} 
Since the orientation of $z^{\prime}$ changes randomly with time, a static field effectively behaves as a field which fluctuates on the timescale of rotational diffusion $1/k_d$ \cite{Liam}. To demonstrate this
consider a static magnetic field with components $B_z$ and $B_x$ along the $z$ and $x$ axes (without loss of generality, $B_y$ = 0). 
Unlike the geometric phase, the phase evolution due to a magnetic field depends explicitly on the azimuthal
angle $\phi$ as well as the polar angle $\theta$ of $z^{\prime}$. The Fokker-Planck equation is then:
\begin{eqnarray}
&\frac{1}{k_d}\frac{d}{dt} p(\t_2,\p,\Phi,t|\t_1) =- \frac{\partial}{\partial \t_2}\left [ \sin \t_2 \frac{\partial}{\partial \t_2}\left ( \frac{p(\t_2,\p,\Phi,t|\t_1)}{\sin\t_2}\right )\right ] \notag  \\ & + \frac{1}{\sin^2\t_2} \left (\frac{\partial}{\partial \p}+\cos\theta_2 \frac{\partial}{\partial \Phi} \right )^2 p(\t_2,\p,\Phi,t|\t_1) \notag \\ & - \left (\frac{\gamma B_z}{k_d} \cos\t_2 + \frac{\gamma B_x}{k_d} \sin\t_2\cos\p \right )\frac{\partial}{\partial \Phi}  p(\t_2,\p,\Phi,t|\t_1). \nonumber \\ \label{FP_2}
\end{eqnarray}
The resulting signal envelope [Eq.~(\ref{envelope})] is plotted in Fig. 5(b) and the decoherence time (defined here as the time taken for the signal envelope to reach $1/e$ of its initial value) is plotted in Fig. 5(c). 

When $\gamma B \gg k_d$, the shape of the dephasing signal is
close to Gaussian, the characteristic shape for slowly fluctuating fields. In this regime the static field effectively fluctuates on the timescales of $1/k_d$ due to the rotation of the crystal. The crossover between Gaussian and exponential decay occurs on the timescale of $1/k_d$, but the exponential behaviour is hidden since the geometric phase accumulation causes dephasing to occur on at least this timescale. The decay is significantly stronger for fields in the $x$ (or $y$) directions than in the $z$ direction. This is because the largest Rabi rotations are experienced by the ensemble members for whom $\theta \approx \pi/2$ which are most sensitive to fields in the $x$-$y$ plane.

For weak fields, the sensitivity approaches zero, which can be seen in the vanishing gradient for small $B$ in Fig. 5(c). To sense small DC fields, a strong static magnetic field could be applied to bring the field sensitivity to within an appropriate range. The nonlinearity of field sensitivity also allows the possibility of vector magnetometry. By applying a DC field in one direction, only DC fields in that direction would be detected, allowing both the magnitude and the direction of DC magnetic fields to be imaged. Given the optimal strength of applied DC field, the magnetic field sensitivity is
\begin{equation}
\Delta B \sqrt T \approx \frac{1}{\gamma \alpha\sqrt{N \tau_d}}, \label{DC_sensitivity}.
\end{equation}
where $1/\tau_d\approx \/T^*_2+k_d$. The solid black curve in Fig. 4 shows the DC magnetic field sensitivity as a function of the crystal radii in room temperature water. 

\subsection{Fluctuating magnetic fields}
Many biological systems display behaviour that is neither static nor periodic, but stochastically fluctuating. Although the details may vary, a fluctuating field can be characterised by a correlation time $t_c$, and a mean squared field strength $\langle B^2 \rangle$ \cite{Liam}. For the tumbling NV system, a magnetic field correlation time larger than the rotational diffusion time will mean that the NV experiences a field that is essentially static over its coherence time, although statistically distributed over many runs. In this case, the effect will be very similar to that for a DC field. 

For the case of a fluctuating field with correlation time $t_c \ll 1/k_d$ and mean
squared field strength $\langle B^2 \rangle = B_{FC}^2 \ll 1/(\gamma t_c)^2$ which is isotropic,  the Fokker-Planck equation is
\begin{eqnarray}& &\frac{1}{k_d} \frac{dp(\t_2,\Phi,t|\t_1)}{dt} =- \frac{\partial}{\partial \t_2}\left [ \sin \t_2 \frac{\partial}{\partial \t_2}\left ( \frac{p\left(\theta_2 , \Phi; t 
|\theta_1\right)}{\sin\t_2}\right )\right ] \nonumber \\
&+& \left(\cot^2\t_2+\frac{\gamma^2 B_{FC}^2t_c}{6k_d}\right) \frac{\partial^2 p\left(\theta_2 , \Phi; t 
|\theta_1\right)}{\partial \Phi^2},
\end{eqnarray}
which is similar to that for the geometric phase alone (\ref{FP}). As such the fluctuating field produces exponential decay of the signal, on a timescale of $1/(\gamma B_{\rm FC} \sqrt{t_c})^2$. The implication of this result is that the decoherence rate is linear in $B_{\rm FC}^2 t_c$.

\subsection{Oscillating magnetic fields}
One application of a nanoscale magnetometer is to detect magnetic resonance from a small number of spins, which can be achieved by measuring the AC field produced by their effective precession, or by driving them via an external microwave field. Applying a spin-echo pulse sequence ($\frac{\pi}{2} - \frac{\tau}{2} - \pi - \frac{\tau}{2} - \frac{\pi}{2}$) effectively rectifies an AC magnetic field of period $2\tau$, allowing its amplitude to be detected while masking fields oscillating at different frequencies. A spin-echo pulse sequence has the additional advantage of dramatically increasing the dephasing time of an NV magnetometer, by refocusing the divergent phases accrued due to slowly fluctuating magnetic fields. 

In detecting magnetic resonance, a strong DC field is usually applied. A tumbling crystal will see such a field as if it were fluctuating on the timescale of the crystal's rotation, causing dephasing on a timescale, $\tau_{n\pi}$, given by:
\begin{equation}
\tau_{n\pi}^3 \approx \frac{(n+1)^2}{4k_B}\left (\frac{1}{\gamma B_{DC}}\right )^2,
\end{equation}
where $n$ is the number of  $\pi$-pulses employed in a concatentated spin-echo pulse sequence.

Taking the limit in which very little rotation occurs on the timescale of free evolution.
The problem is then a matter of simply averaging over the possible crystal orientations
and initial phases of the AC field, rather than solving a Fokker-Planck equation. Consider
a spin-echo pulse sequence during which the crystal is oriented at some $\theta$ and the free evolution
time $\tau$ matches the AC fieldÕs period. The final population of $\psi_{z^{\prime}}^{(0)}$ is given by
\begin{eqnarray}
P_0(\theta,\Phi_{\rm field}) &=&
\frac{1}{2}+
\frac{1}{2}\cos^2 \Theta_1 \cos^2 \Theta_2  \nonumber \\
&\pm&
\frac{1}{4} \sin^2 \Theta_1(1-\cos^2 \Theta_2) \cos(2\Phi_{\rm AC}) \nonumber \\
\end{eqnarray}
where $\Theta_1 = a(\pi/2) \sin \theta$ and $\Theta_2 = a\pi \sin \theta$ are the angles through which the spin is rotated
during the $\pi/2$ pulses and $\pi$-pulse respectively. For an AC field oscillating along the $z$-axis, the phase acquired by the NV centre before the $\pi$ pulse is $\Phi_{\rm AC} = (\tau/2)\gamma B_{\rm AC}\cos\theta \cos \Phi_{\rm field}$,
where $\Phi_{\rm field}$ is the phase of the AC field at the start of the period, and the phase acquired
after the $\pi$ pulse is $-\Phi_{\rm AC}$. 

In the limit that the strength of the AC field to be detected is much smaller than the applied DC field strength, it can be assumed that the crystal is stationary over the timescale of free evolution, although an average over the full ensemble of possible orientations still must be performed. To find the ensemble-averaged signal, we integrate over all phases of the AC field, $\Phi_{\rm field}$, and all orientations, $\theta$, weighted by $\frac{1}{2} \sin \theta$:
\begin{eqnarray}
S =\int_0^{\pi}d \theta
\frac{\sin\theta}{2}
\int_0^{2\pi}
d\Phi_{\rm field}\frac{P_0(\theta,\Phi_{\rm field})}{2\pi}.
\end{eqnarray}
Figure 5(d) shows the signal envelope due to an AC field. As with DC fields, the optimum AC field strength is such that the dephasing due to the AC field matches the dephasing rate from other sources. The optimal sensitivity to an AC field is the same as that given in (\ref{DC_sensitivity}), except with $1/\tau_d \approx 1/T_2 + 1/\tau_{n\pi} + k_d$.
Using a concatenated pulse sequence can increase the decoherence time. Concatenation also reduces the bandwidth of AC field detection, which is important if different AC frequency components are to be distinguished.

\section{Conclusion}

We have shown that a freely diffusing diamond nanocrystal can act as a sensitive magnetometer with sensitivity comparable to that of a fixed crystal magnetometer. We have also shown that, through the accumulation of a geometric phase, the fluorescence signal contains information about the crystal's rotational diffusion rate. This information can be gathered without applying any microwave control sequences, and is thus achievable with little modification to the existing experimental setups that use fluorescent colloidal nanocrystals as biological markers.


Before the tumbling nanocrystal system can be used in biological applications, further theoretical and experimental work will be required. It will help to understand anisotropic rotation, and to investigate the effect of the fields produced by a specific biological process rather than the idealised fields considered here. The most challenging experimental task will be finding protocols to either track or control rapid spatial movement of the crystal, while still achieving satisfactory microwave control of the quantum system within \cite{Scholten}.

\section{Acknowledgements} 
This research was supported by the Australian Research Council Centre of Excellence for Quantum Computation and Communication Technology (CE110001027). L.C.L.H. was supported under an Australian Research Council Professorial Fellowship (DP0770715).

\section{Appendix A}
\subsection{Ensemble average population mixing rate due to non-adiabatic rotation}
To evaluate the properties of an ensemble measurement consider an arbitrary initial state of the NV centre, where each state has a probability weighting of $P_{m}$ and phase $\Phi_m$. The populations of the spin sublevels after some time interval $\delta t$ is:
\begin{eqnarray}
P_1^{\prime}&=&P_1-k\delta t P_1+k \delta t P_0 \nonumber \\&+&2\sqrt{P_1P_0}\sqrt{k \delta t}\cos\left(\Phi_0-\Phi_1+\delta \phi\right) \\
P_0^{\prime}&=&P_0-2k \delta t P_0+k\delta t P_1+k \delta t P_{-1} \nonumber \\&-&2\sqrt{P_1P_0}\sqrt{k \delta t }\cos\left(\Phi_0-\Phi_1+\delta \phi\right)\nonumber \\
&-&2\sqrt{P_{-1}P_0}\sqrt{k \delta t}\cos\left(\Phi_0-\Phi_{-1}+\delta \phi\right)\\
P_{-1}^{\prime}&=&P_{-1}-k\delta t P_{-1}+k \delta t P_0 \nonumber \\&+&2\sqrt{P_{-1}P_0}\sqrt{k \delta t}\cos\left(\Phi_0-\Phi_{-1}-\delta \phi\right),
\end{eqnarray}
where
\begin{eqnarray}
k=\frac{\delta \theta^2}{D^2\delta t^3}\left[1-\cos (\delta tD\right)].
\end{eqnarray}
Modeling the crystal's Brownian rotation as a random walk of small uncorrelated  angular steps of duration $\delta t$ and angular change $\delta \theta$ on timescales much longer than $\delta t$ the ensemble average of ${\dot P}_{\alpha}$ can be determined:
\begin{eqnarray}
\frac{d\langle P_1\rangle}{dt}&=&-\langle k \rangle \langle P_1 \rangle+\langle k \rangle \langle P_{-1} \rangle \\
\frac{d\langle P_0\rangle}{dt}&=&\langle k \rangle \langle P_1 \rangle+\langle k \rangle \langle P_{-1} \rangle-2\langle k\rangle\langle P_0 \rangle \\
\frac{d\langle P_{-1}\rangle}{dt}&=&-\langle k \rangle \langle P_{-1} \rangle+\langle k \rangle \langle P_{1} \rangle ,
\end{eqnarray}
To find the ensemble average of $k$ we need to know how $\delta t$ is distributed. If we multiply
the Langevin equation (\ref{langevin}) by ${\bm \omega} (0)$ and take the ensemble average we find that the correlation function of the angular velocity decays exponentially over time,
\begin{eqnarray}
\langle{\bm \omega}(t)\cdot {\bm \omega}(0) \rangle = \langle{\bm \omega}(0)\cdot {\bm \omega}(0) \rangle e^{-t/t_d},
\end{eqnarray}
where we have defined the damping time, $t_d=I/\gamma_d$. In order to retain this correlation
function in our discrete {\it random walk} approximation, the duration of a given step, $\delta t$, must be exponentially distributed. At any given instant, the fraction of the ensemble that
is experiencing a step of duration $\delta t$ is then given by the probability density:
\begin{eqnarray}
p(\delta t)=\frac{\delta t}{t^2_d}e^{-\delta t/t_d}.
\end{eqnarray}
We do not need to know the exact distribution of angular velocities, merely that $\langle (\delta \theta)^2 / (\delta t)^2\rangle=2/3\langle {\bm \omega} \cdot {\bm \omega} \rangle =2k_d/t_d$. The ensemble average population mixing rate is then given by
\begin{eqnarray}
\langle k \rangle&=&\frac{1}{D^2}\left\langle\frac{\delta \theta^2 }{\delta t^2}\right\rangle\left\langle \frac{1}{\delta t}\left(1- \cos (\delta t D)\right)\right\rangle  \\
&=&\frac{2k_d}{1+t_d^2D^2}.
\end{eqnarray}

\end{document}